\RequirePackage[OT1]{fontenc}
\documentclass[letterpaper, 10 pt, conference]{ieeeconf} 

\IEEEoverridecommandlockouts  
\usepackage{amsmath}
\usepackage{amssymb}
\usepackage{amsthm}
\usepackage{mathtools}
\usepackage{derivative}
\usepackage{bm}

\usepackage[noadjust]{cite} 

\usepackage{graphicx}
\usepackage[export]{adjustbox} 
\graphicspath{{figures/}} 

\newtheorem{theorem}{Theorem}
\newtheorem{lemma}{Lemma}
\newtheorem{proposition}{Proposition}

\theoremstyle{definition}
\newtheorem{definition}{Definition}

\newtheorem{remark}{Remark}

\newtheorem{example}{Example}

\newcommand{\argmin}{\operatornamewithlimits{arg\,min}}

\newcommand{\hcat}{\operatorname{hcat}}

\newcommand{\R}{\mathbb{R}}

\newcommand{\C}{\mathcal{C}}

\newcommand{\T}{^\top}

\newcommand{\bzero}{\mathbf{0}}

\newcommand{\bb}{\mathbf{b}}
\newcommand{\bc}{\mathbf{c}}
\newcommand{\bd}{\mathbf{d}}

\renewcommand{\bf}{\mathbf{f}} 
\newcommand{\bg}{\mathbf{g}}

\newcommand{\bk}{\mathbf{k}}

\newcommand{\bu}{\mathbf{u}}

\newcommand{\bx}{\mathbf{x}}

\newcommand{\bz}{\mathbf{z}}

\newcommand{\bA}{\mathbf{A}}
\newcommand{\bB}{\mathbf{B}}

\newcommand{\bF}{\mathbf{F}}
\newcommand{\bG}{\mathbf{G}}

\newcommand{\bI}{\mathbf{I}}

\newcommand{\bK}{\mathbf{K}}

\newcommand{\bP}{\mathbf{P}}
\newcommand{\bQ}{\mathbf{Q}}

\newcommand{\bW}{\mathbf{W}}

\newcommand{\bY}{\mathbf{Y}}

\newcommand{\btheta}{\bm{\theta}}
\newcommand{\btau}{\bm{\tau}}
\newcommand{\bphi}{\bm{\phi}}

\newcommand{\bGamma}{\bm{\Gamma}}

\newcommand{\bPhi}{\bm{\Phi}}

\newcommand{\bhat}[1]{\skew{3}{\hat}{\bm{#1}}}
\newcommand{\bdothat}[1]{\skew{3}{\dot}{\skew{3}{\hat}{\bm{#1}}}}
\newcommand{\btilde}[1]{\skew{3}{\tilde}{\bm{#1}}}

\makeatletter
\let\NAT@parse\undefined
\makeatother

\usepackage{xcolor}
\definecolor{myblue}{RGB}{49, 114, 174}
\definecolor{myred}{rgb}{0.796, 0.235, 0.2}
\definecolor{mygreen}{rgb}{0.22, 0.596, 0.149}
\definecolor{mypurple}{rgb}{0.584,0.345,0.698}

\usepackage{hyperref}
\hypersetup{
    colorlinks=true,
    linkcolor=black,
    filecolor=magenta,
    citecolor=black,      
    urlcolor=cyan,
    pdftitle={title},
}

\title{\textbf{ Uncertainty Quantification for Recursive Estimation \\ in Adaptive Safety-Critical Control}}

\author{Max H. Cohen$^1$, %
Makai Mann$^2$, %
Kevin Leahy$^3$, %
and Calin Belta$^4$ %
\thanks{$^1$ Department of Mechanical and Civil Engineering,
California Institute of Technology, Pasadena, CA 91125; \texttt{maxcohen@caltech.edu}}
\thanks{$^2$ Massachusetts Institute of Technology Lincoln Laboratory, Lexington, MA 02421; \texttt{makai.mann}\texttt{@ll.mit.edu}}
\thanks{$^3$ Department of Robotics Engineering, Worcester Polytechnic Institute, Worcester, MA 01609; \texttt{kleahy}\texttt{@wpi.edu}}
\thanks{$^4$ Department of Electrical and Computer Engineering, University of Maryland, College Park, MD 20742; \texttt{calin@umd.edu}}
\thanks{This work is supported by the NSF under grants DGE-1840990
and IIS-2024606. Any opinions, findings, conclusions or recommendations expressed in this material are those of the author(s) and do not necessarily reflect the views of the NSF. The NASA University Leadership initiative (grant \#80NSSC20M0163) provided funds to assist the authors with their research, but this article solely reflects the opinions and conclusions of its authors and not any NASA entity.}
}

\begin{document}
    \maketitle 

    \begin{abstract}
        In this paper, we present a framework for online parameter estimation and uncertainty quantification in the context of adaptive safety-critical control. The key insight enabling our approach is that the parameter estimate generated by the continuous-time recursive least squares (RLS) algorithm at any point in time is an affine transformation of the initial parameter estimate. This property allows for parameterizing such estimates using objects that are closed under affine transformation, such as zonotopes, and enables the efficient propagation of such set-based estimates as time progresses. We illustrate how such an approach facilitates the synthesis of safety-critical controllers for systems with parametric uncertainty and additive disturbances using control barrier functions, and demonstrate the utility of our approach through illustrative examples.
    \end{abstract}

    \section{Introduction}
    As autonomous systems are deployed in complex environments, it is essential for such systems to \emph{adapt} to uncertainties that may be challenging to characterize until deployment. Recently, there has been a surge of interest in using data-driven and learning-based techniques to endow such systems with adaptive capabilities. This integration of learning and control, however, raises questions regarding the reliability and safety of such learning-enabled systems. In a control-theoretic context, simultaneous learning and control is the focus of \emph{adaptive control} \cite{IoannouSun,Krstic,Cohen} where the primary objective is stabilization of dynamical systems with parametric uncertainties. 

    Although adaptive control schemes with stability guarantees have existed for decades, there has been much less work on extending such ideas to address other important properties, such as \emph{safety}. The property of safety -- informally thought of as requiring a system to ``never do anything bad" -- has often been formalized using the notion of set invariance \cite{AmesECC19}. In an analogous fashion to how Lyapunov functions are leveraged to certify the stability of equilibrium points without integrating a system's vector fields, the concept of a \emph{barrier function} \cite{AmesTAC17} provides a methodology to certify set invariance without computing a system's reachable set. Such ideas naturally extend to control systems via control barrier functions (CBFs) \cite{AmesECC19,AmesTAC17} that allow for the design of controllers that enforce set invariance.
    
    Given the duality between Lyapunov and barrier functions \cite{AmesECC19}, recent works have begun to transfer Lyapunov-based adaptive control techniques to a barrier function setting. For example, \cite{TaylorACC20} introduced the notion of an adaptive CBF (aCBF) that allows for synthesizing controllers and parameter update laws enforcing set invariance conditions for nonlinear systems with uncertain parameters. A key insight of follow-up works \cite{Cohen,LopezLCSS21,DixonACC21,CohenACC22,CohenACC23,LopezACC23,PanagouECC21,PanagouArXiV22} is that directly extending classical Lyapunov-based techniques to a barrier function setting can lead to overly conservative controllers, and that such conservatism can be reduced by taking a \emph{robust} adaptive control approach. Here, one assumes known bounds on the system parameters, designs a robust controller enforcing safety for all possible realizations of the parameters, and then reduces the robustness of this controller as the uncertainty in the parameter estimates shrinks. 
    
    The success of the aforementioned robust adaptive approaches relies on various \emph{uncertainty quantification} mechanisms that associate an uncertainty metric to a given parameter estimate. For aCBFs, such an idea was introduced in \cite{LopezLCSS21}, where uncertainty quantification is achieved using \emph{set membership identification} (SMID) -- a technique commonly used in adaptive model predictive control \cite{MorariAutomatica14,GuaySCL14,CannonIFAC16} that maintains a set-based parameter estimate, rather than a pointwise estimate as in classical adaptive control \cite{IoannouSun}. Such set-based estimates are generally computed by solving an optimization problem at each time instant and often allow one to account for additive disturbances in addition to uncertain parameters \cite{CannonIFAC16}. One limitation of SMID approaches is that the number of constraints in the optimization problem may grow unbounded for extended time horizons, requiring various complexity reduction techniques to facilitate real-time implementation \cite{MorariAutomatica14}. Alternate uncertainty quantification techniques used for aCBFs include concurrent learning \cite{Chowdhary,ChowdharyACC11} -- an adaptive control method that facilitates parameter convergence under relaxed excitation conditions. Such conditions are related to the rank of a data matrix that is updated as time progresses, and allows for bounding the rate of convergence of the parameter estimates using Lyapunov-like arguments \cite{DixonACC21,CohenACC22,CohenACC23,Cohen}. Such an approach allows for efficiently computing bounds on the parameter estimation error; however, the bounds are often conservative, especially in the presence of additive disturbances.

    In this paper, we unite SMID and concurrent learning adaptive control for online parameter estimation and uncertainty quantification in the context of adaptive safety-critical control. Our starting point is the classic \emph{recursive least squares} (RLS) algorithm for online parameter estimation. Inspired by \cite{Chowdhary}, we first demonstrate how incorporating a history stack of data into RLS facilitates parameter convergence under milder conditions than the traditional PE condition, and analyze the impact of additive disturbances on such convergence results (Sec. \ref{sec:RLS}). We then demonstrate how the pointwise parameter estimates produced by RLS can be directly used to construct a set-based parameter estimate, allowing one to quantify the uncertainty in the original pointwise estimates. Our key insight is that the parameter estimate produced by RLS at any given point in time is an affine transformation of the initial estimate (Sec. \ref{sec:UQ}). This observation allows for parameterizing such estimates using various objects that are closed under affine transformation, such as zonotopes \cite{GirardHSCC05}. By replacing the initial pointwise parameter estimate with a zonotope containing the true parameters, this approach allows to compute, at each instant in time, another zonotope containing the true parameters using only the initial zonotope and pointwise estimate produced by RLS. Ultimately, this results in a \emph{recursive} SMID technique that maintains the computational efficiently of RLS while producing set-based parameter estimates that allow for uncertainty quantification in the presence of additive disturbances. We demonstrate how this approach facilitates the synthesis of safety-critical controllers for uncertain nonlinear systems (Sec. \ref{sec:cbf}) and illustrate the efficacy of our approach with various examples.

    \section{Preliminaries and Problem Formulation}\label{sec:prelim}
    \noindent \textit{Notation}: For a matrix $\bA\in\R^{n\times n}$ the notation $\bA\geq 0$ ($\bA >0$) denotes that $\bA$ is positive semidefinite (positive definite). The symbol $\bI$ denotes the identity matrix. Given matrices $\bA,\bB$ with the same number of rows, we define $\hcat(\bA,\bB)$ as their horizontal concatenation. We use $\|\cdot\|_p$ to denote the (induced) $p$-norm for a (matrix) vector. For a continuously differentiable function $h\,:\,\R^n\rightarrow\R$ and a vector field $\bf\,:\,\R^n\rightarrow\R^n$, we define $L_{\bf}h(\bx)\coloneqq \nabla h(\bx)\cdot\bf(\bx)$ as the Lie derivative of $h$ along $\bf$.
    
    The high-level problem considered in this paper is online parameter estimation. Here, we construct an online estimate $\bhat{\theta}\in\R^p$ of a fixed parameter vector $\btheta\in\R^p$ -- typically corresponding to uncertain parameters of a control system -- based upon input-output data generated by a parameter measurement model. Keeping in line with standard approaches in adaptive control \cite{IoannouSun,Krstic}, we assume this measurement model is affine in the parameters and of the form:
    \begin{equation}\label{eq:LRE}
        y(t) = \bphi(t)\T\btheta + w(t),
    \end{equation}
    where the mappings\footnote{We assume $y(t)$ is a scalar only for ease of exposition - the results here can be extended to vector-valued $y(t)$ and matrix-valued $\bphi(t)$.} $y\,:\,\R_{\geq0}\rightarrow\R$, $\bphi\,:\,\R_{\geq0}\rightarrow\R^p$, and $w\,:\,\R_{\geq0}\rightarrow\R$ associate to each $t\geq0$ a target value $y(t)\in\R$, feature vector $\bphi(t)\in\R^p$, and a disturbance input $w(t)\in\R$. Throughout this paper, we make the standing assumption that all signals in \eqref{eq:LRE} are bounded. The primary objective in online parameter estimation is to build an estimate $\bhat{\theta}\,:\,\R_{\geq0}\rightarrow\R^p$ of $\btheta$ completely online based upon knowledge of the signals $y(\cdot)$ and $\bphi(\cdot)$. Most often, this is accomplished by solving the optimization problem:
    \begin{equation}\label{eq:parameter-optimization}
        \min_{\bhat{\theta}\in\R^p}J(\bhat{\theta},t),
    \end{equation}
    where $J\,:\,\R^p\times\R_{\geq0}\rightarrow\R$ is a cost function, continuously differentiable in its first argument and piecewise continuous in its second, that represents a metric between observed measurements $y(t)$ and predicted measurements $\bphi(t)\T\bhat{\theta}$, allowing to construct the parameter estimate at time $t$ as:
    \begin{equation}\label{eq:parameter-estimate}
        \bhat{\theta}(t) = \argmin_{\bhat{\theta}\in\R^p} J(\bhat{\theta},t).
    \end{equation}
    Rather than explicitly solving \eqref{eq:parameter-optimization} for every $t$, one may recursively implement the solution \eqref{eq:parameter-estimate} to optimization problem \eqref{eq:parameter-optimization} by propagating the ordinary differential equation $\bdothat{\theta} = \btau(\bhat{\theta},t)$, where $\btau\,:\,\R^p\times\R_{\geq0}\rightarrow\R^p$ is a vector field, locally Lipschitz in its first argument and piecewise continuous in its second, from an initial condition $\bhat{\theta}_0\in\R^p$ corresponding to the best guess of $\btheta$. This allows for efficiently constructing a parameter estimate $\bhat{\theta}(t)$ for each $t$ from streaming data, but generally does not provide information related to the uncertainty of such an estimate. In this paper, we seek to develop an efficient parameter estimation routine that also allows for quantifying the uncertainty in such an estimate, which we accomplish by recursively updating an entire \emph{set} of possible parameters consistent with the observed data.

    \section{Recursive-Batch Least Squares Regression}\label{sec:RLS}
    \begin{figure}
        \centering
        \vspace{0.2cm}
        \includegraphics{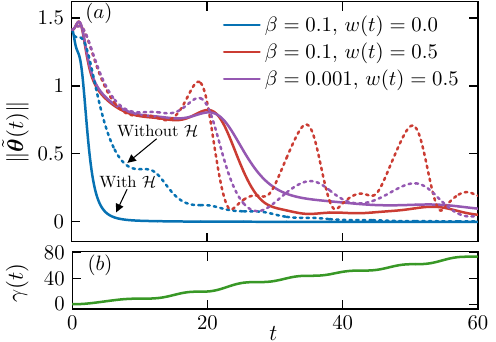}
        \vspace{-0.4cm}
        \caption{Evolution of the parameter estimation errors (a) and value of $\gamma$ (b) from \eqref{eq:FE} corresponding to Example \ref{example:simple-example}. The solid curves correspond to the estimates using $\mathcal{H}$ and the dashed curves to those that do not exploit $\mathcal{H}$.}
        \label{fig:simple-example}
        \vspace{-0.6cm}
    \end{figure}
    In this section, we present a hybrid approach that combines aspects of batch and recursive least squares techniques for regression in the context of estimating the parameters $\btheta\in\R^p$ of the regression model from \eqref{eq:LRE} based on ideas from concurrent learning adaptive control \cite{Chowdhary,DixonIJACSP19,Cohen}. Central to such approaches is the concept of a \emph{history stack}.
    \begin{definition}
        A collection $\mathcal{H}(t) = \{(y(t_j),\bphi(t_j)\}_{j=1}^M$, where $y$ and $\bphi$ are as in \eqref{eq:LRE} and $t_j\in\R_{\geq0}$, is said to be a \emph{history stack} for \eqref{eq:LRE}.
    \end{definition}
    One may think of a history stack $\mathcal{H}$ simply as a dataset of $M\in\mathbb{N}$ target and feature values $(y(t_j),\bphi(t_j))\in\R\times\R^p$ sampled from \eqref{eq:LRE} at various point in time $t_j\in\R_{\geq0}$. Note that the stack itself may vary with time as data is added or removed. Thus, $\mathcal{H}(t)$ returns the data present in the stack at time $t$, which need not be the same as the data present in the stack at time $t'$ in the sense that we may have $\mathcal{H}(t)\neq\mathcal{H}(t')$ for $t\neq t'$.
    Algorithms for populating $\mathcal{H}$ with data are outlined in \cite{ChowdharyACC11}. As originally shown in \cite{Chowdhary}, constructing a parameter update law using the data present in $\mathcal{H}$, rather than only the data generated by \eqref{eq:LRE} at each time, allows for relaxing the classical persistence of excitation (PE) condition often required for parameter convergence. To introduce these relaxed conditions, we concatenate the data in $\mathcal{H}$ as:
    \begin{equation}\label{eq:Y-Phi}
        \bY(t) \coloneqq 
        \begin{bmatrix}
            y(t_1) \\ y(t_2) \\ \vdots \\ y(t_M)
        \end{bmatrix}\in\R^M,
        \quad
        \bPhi(t) \coloneqq\begin{bmatrix}
            \bphi(t_1)\T \\ \bphi(t_2)\T \\ \vdots \\ \bphi(t_M)\T
        \end{bmatrix}\in\R^{M\times p},
    \end{equation}
    which satisfy:
    \begin{equation}\label{eq:LRE-2}
        \bY(t) = \bPhi(t)\btheta + \bW(t),
    \end{equation}
    where $\bW(t)\coloneqq(w(t_1),w(t_2),\dots,w(t_M))\in\R^M$. The following definition outlines a relaxed excitation condition for parameter convergence.
    \begin{definition}
        A history stack $\mathcal{H}$ is said to satisfy the finite excitation (FE) condition if there exists a time $T\in\R_{\geq0}$ and positive constant $\gamma\in\R_{>0}$ such that for all $t\in[T,\infty)$
        \begin{equation}\label{eq:FE}
            \bPhi(t)\T\bPhi(t) \geq \gamma \bI.
        \end{equation}
    \end{definition}
    The FE condition\footnote{In contrast, the PE condition requires the existence of a time $T>0$ such that the matrix $\int_{t}^{t+T}\bPhi(s)\T\bPhi(s)\mathrm{d}s$ is invertible for all $t\geq0$.} requires $\bPhi(t)\T\bPhi(t)$ to be invertible\footnote{The data collection algorithms in \cite{ChowdharyACC11} ensure that $\lambda_{\min}(\bPhi(t)\T\bPhi(t))$ is non-decreasing; hence, if \eqref{eq:FE} holds at a single instant in time, then it also holds for all times thereafter.} for all $t\geq T$, thereby ensuring the existence of a solution to $\bY(t)=\bPhi(t)\btheta$ for each $t\geq T$. A necessary condition for such a solution to exist is $M\geq p$. To construct a parameter update law using $\mathcal{H}$, we introduce the cost function:
    \begin{equation}\label{eq:LS-cost}
    \begin{aligned}
        J(\bhat{\theta},t) = & \frac{1}{2}\int_{0}^{t}e^{-\beta(t-s)}\| \bY(s) - \bPhi(s)\bhat{\theta}\|_2^2\mathrm{d}s \\ &+ \frac{1}{2}e^{-\beta t}(\bhat{\theta} - \bhat{\theta}_0)\T \bP(\bhat{\theta} - \bhat{\theta}_0),
    \end{aligned}
    \end{equation}
    where $\bP\in\R^{p\times p}$ is positive definite and $\beta\in\R_{>0}$ is a discount factor. The estimate that minimizes the above expression is given by:
    \begin{equation}\label{eq:theta-hat}
        \bhat{\theta}(t) = \bGamma(t)\bigg[e^{-\beta t}\bP\bhat{\theta}_0 + \int_{0}^t e^{-\beta(t-s)} \bPhi(s)\T \bY(s)\mathrm{d}s\bigg],
    \end{equation}
    \begin{equation}\label{eq:Gamma}
        \bGamma(t) = \bigg[e^{-\beta t}\bP +  \int_{0}^{t}e^{-\beta (t-s)}\bPhi(s)\T\bPhi(s)\mathrm{d}s\bigg]^{-1},
    \end{equation}
    which can be recursively computing by solving:
    \begin{equation}\label{eq:theta-hat-dot}
        \bdothat{\theta} = \bGamma\bPhi(t)\T\Big[\bY(t) - \bPhi(t)\bhat{\theta}\Big],
    \end{equation}
    \begin{equation}\label{eq:Gamma-dot}
        \dot{\bGamma} = \beta\bGamma -\bGamma \bPhi(t)\T\bPhi(t)\bGamma,
    \end{equation}
    with initial conditions $\bhat{\theta}(0)=\bhat{\theta}_0$ and $\bGamma(0)=\bP^{-1}$. The above ODEs correspond to the continuous-time recursive least-squares (RLS) algorithm, which can be interpreted as a Kalman filter for the static dynamical system $\dot{\btheta}=0$ with measurement $\bY(t)=\bPhi(t)\btheta + \bW(t)$, where $\bGamma$ corresponds to the covariance\footnote{Since $\btheta$ is not viewed as a random variable, formally, $\bGamma$ is not a covariance matrix; however, as demonstrated later on $\bGamma$ does indeed encode a measure of uncertainty associated with the estimate $\bhat{\theta}$.} of the estimate $\bhat{\theta}$.
    By following the same steps as in the proof of \cite[Corollary 4.3.2]{IoannouSun}, one can show that, provided the FE condition is satisfied, then there exist positive constants $\gamma_1,\gamma_2\in\R_{>0}$ such that:
    \begin{equation}\label{eq:Gamma-bounds}
        \gamma_1 \bI \leq \bGamma(t) \leq \gamma_2\bI,\quad \forall t\in\R_{\geq0},
    \end{equation}
    where $\gamma_2$ is inversely proportional to $\gamma$ in \eqref{eq:FE}. This implies $\bGamma(t)$ is bounded and positive definite for all $t\in\R_{\geq0}$, thereby implying the existence of $\bGamma(t)^{-1}$ for all $t\in\R_{\geq0}$. Our objective now is to quantify the uncertainty in the estimate $\bhat{\theta}$, which is accomplished by studying the evolution of the parameter estimation error:
    \begin{equation}\label{eq:theta-tilde}
        \btilde{\theta}(t) \coloneqq \btheta - \bhat{\theta}.
    \end{equation}
    The following lemma provides a useful representation of $\btilde{\theta}$.
    \begin{lemma}\label{lemma:theta-tilde}
        Let $\mathcal{H}$ satisfy the FE condition and suppose that $\bhat{\theta}$ and $\bGamma$ satisfy \eqref{eq:theta-hat} and \eqref{eq:Gamma}, respectively. Then, for all $t\in\R_{\geq0}$, the parameter estimation error \eqref{eq:theta-tilde} satisfies:
        \begin{equation}\label{eq:theta-tilde-RLS}
        \begin{aligned}
            \btilde{\theta}(t) = & e^{-\beta t}\bGamma(t)\bP(\btheta - \bhat{\theta}_0) \\
             &- \bGamma(t)\int_{0}^{t}e^{-\beta(t-s)}\bPhi(s)\T\bW(s)\mathrm{d}s.
        \end{aligned}
        \end{equation}
    \end{lemma}
    \begin{proof}
        Using \eqref{eq:LRE-2} and \eqref{eq:theta-hat}, $\bhat{\theta}$ can be expressed as:
        \begin{equation*}
        \begin{aligned}
            \bhat{\theta}(t) = & e^{-\beta t}\bGamma(t)\bP\bhat{\theta}_0 + \bGamma(t)\int_{0}^{t}e^{-\beta(t-s)}\bPhi(s)\T\bPhi(s)\btheta\mathrm{d}s \\
            & + \bGamma(t)\int_{0}^{t}e^{-\beta(t-s)}\bPhi(s)\T\bW(s)\mathrm{d}s.
        \end{aligned}
        \end{equation*}
        Using the fact that $\bGamma(t)$ is invertible for each $t\in\R_{\geq0}$, we have $\btheta = \bGamma(t)\bGamma(t)^{-1}\btheta$, which implies that
        \begin{equation*}
            \begin{aligned}
                \btheta = e^{-\beta t}\bGamma(t)\bP\btheta + \bGamma(t)\int_{0}^{t}e^{-\beta (t-s)}\bPhi(s)\T\bPhi(s)\btheta\mathrm{d}s.
            \end{aligned}
        \end{equation*}
        Substituting the previous two expressions into \eqref{eq:theta-tilde} then yields \eqref{eq:theta-tilde-RLS}, as desired.
    \end{proof}
    In the disturbance-free case (i.e., $\bW(t)\equiv 0$), the preceding lemma shows that $\bhat{\theta}$ exponentially converges to $\btheta$ when $\mathcal{H}$ satisfies the FE condition. The following result provides a conservative bound on $\btilde{\theta}$ when disturbances are present.

    \begin{proposition}\label{proposition:theta-tilde}
        Let the conditions of Lemma \ref{lemma:theta-tilde} hold and define 
        \begin{equation}\label{eq:w-bar}
            \bar{w}\coloneqq \sup_{t\geq0}\|\bPhi(t)\T\bW(t)\|_2.
        \end{equation}
        Then, for all $t\geq0$, the parameter estimation error satisfies:
        \begin{equation}\label{eq:theta-tilde-bound}
             \|\btilde{\theta}(t)\|_2 \leq \gamma_2 e^{-\beta t}\|\bP(\bhat{\theta}_0 - \btheta)\|_2 + \frac{\gamma_2\bar{w}}{\beta}.
        \end{equation}
    \end{proposition}
    \begin{proof}
        We begin by noting that:
        \begin{equation*}
            \begin{aligned}
                \|\btilde{\theta}(t)\|_2 = & \bigg\Vert e^{-\beta t}\bGamma(t)\bP(\btheta - \bhat{\theta}_0) \\ 
                & - \bGamma(t)\int_{0}^{t}e^{-\beta(t-s)}\bPhi(s)\T\bW(s)\mathrm{d}s \bigg\Vert_2 \\
                = & \bigg\Vert e^{-\beta t}\bGamma(t)\bP(\bhat{\theta}_0 - \btheta) \\ 
                & + \bGamma(t)\int_{0}^{t}e^{-\beta(t-s)}\bPhi(s)\T\bW(s)\mathrm{d}s \bigg\Vert_2, \\
            \end{aligned}
        \end{equation*}
        which can be upper-bounded as:
        \begin{equation*}
            \begin{aligned}
                \|\btilde{\theta}(t)\|_2 \leq & \|e^{-\beta t}\bGamma(t)\bP(\bhat{\theta}_0 - \btheta) \|_2 \\ 
                & + \bigg\Vert \bGamma(t)\int_{0}^{t}e^{-\beta(t-s)}\bPhi(s)\T\bW(s)\mathrm{d}s \bigg\Vert_2 \\
                \leq & e^{-\beta t}\|\bGamma(t)\bP(\bhat{\theta}_0 - \btheta)\|_2 \\
                & + \bigg\Vert \bGamma(t)\int_{0}^{t}e^{-\beta(t-s)}\bPhi(s)\T\bW(s)\mathrm{d}s \bigg\Vert_2, \\
            \end{aligned}
        \end{equation*}
        where the second equality follows from the homogeneity of norms, the first inequality from the triangle inequality, and the second inequality from the homogeneity of norms. It then follows from \eqref{eq:Gamma-bounds}, the Cauchy-Schwarz inequality, and the fact that the induced matrix 2-norm is monotonic that:
        \begin{equation*}
            \begin{aligned}
                \|\bGamma(t)\bP(\bhat{\theta}_0 - \btheta)\|_2 \leq & \|\bGamma(t)\|_2\cdot\| \bP(\bhat{\theta}_0 - \btheta)\|_2 \\
                \leq & \|\gamma_2\bI\|_2 \cdot\| \bP(\bhat{\theta}_0 - \btheta)\|_2 \\
                \leq & \gamma_2 \| \bP(\bhat{\theta}_0 - \btheta)\|_2.
            \end{aligned}
        \end{equation*}
        Using the Cauchy-Schwarz inequality, we also have:
        \begin{equation*}
            \begin{aligned}
                \eta(t) = & \bigg\Vert \bGamma(t)\int_{0}^{t}e^{-\beta(t-s)}\bPhi(s)\T\bW(s)\mathrm{d}s \bigg\Vert_2 \\ 
                \leq & \|\bGamma(t)\|_2 \cdot \bigg\Vert \int_{0}^{t}e^{-\beta(t-s)}\bPhi(s)\T\bW(s)\mathrm{d}s \bigg\Vert_2. \\ 
            \end{aligned}
        \end{equation*}
        After using \eqref{eq:Gamma-bounds} and the fact that the norm of an integral is upper-bounded by the integral of the norm, the above may be bounded as:
        \begin{equation*}
            \begin{aligned}
                \eta(t) \leq & \gamma_2 \int_{0}^{t} \Big\Vert e^{-\beta(t-s)}\bPhi(s)\T\bW(s) \Big\Vert_2\mathrm{d}s \\
                \leq & \gamma_2 \int_{0}^{t} e^{-\beta(t-s)}\Big\Vert \bPhi(s)\T\bW(s) \Big\Vert_2\mathrm{d}s. \\
            \end{aligned}
        \end{equation*}
        Using \eqref{eq:w-bar} then allows the above to be bounded as:
        \begin{equation*}
            \begin{aligned}
                \eta(t) \leq & \gamma_2\bar{w} \int_{0}^{t} e^{-\beta(t-s)}\mathrm{d}s
                = \frac{\gamma_2\bar{w}}{\beta}\Big(1 - e^{-\beta t}\Big)
                \leq \frac{\gamma_2 \bar{w}}{\beta}.
            \end{aligned}
        \end{equation*}
        Combining the preceding inequalities yields \eqref{eq:theta-tilde-bound}, as desired.
    \end{proof}
    The preceding result provides a conservative bound \eqref{eq:theta-tilde-bound} on the parameter estimation error in the presence of disturbances, and shows that, under the FE condition, the parameter estimates converge to a ball of radius $\tfrac{\gamma_2\bar{w}}{\beta}$ centered at $\btheta$. Recalling that $\gamma_2$ is inversely proportional to $\gamma$ from \eqref{eq:FE}, this implies that the effects of disturbances can be mitigated by increasing\footnote{Although \eqref{eq:theta-tilde-bound} suggests increasing $\beta$ may help attenuate disturbances as well, $\gamma_2$ also depends on $\beta$, and increasing $\beta$ may also decrease $\gamma_2$ (see \cite[Cor. 4.3.2]{IoannouSun} for more details).} $\gamma$ (i.e., by collecting ``better" data -- see \cite{ChowdharyACC11} for data collection strategies). Although the bound in \eqref{eq:theta-tilde-bound} is theoretically appealing, it is challenging to exploit \emph{online} for quantifying the uncertainty in the estimate $\bhat{\theta}$ since it depends on $\gamma_2$, which, in turn, depends on $\gamma$, which is unknown a priori. The main ideas introduced in this section are illustrated in the following simple example.

    \begin{example}\label{example:simple-example}
        Consider the parameter measurement model in \eqref{eq:LRE} defined by $\bphi(t)=[2\sin(t/5)\,\tfrac{1}{2}\cos(2t/5)]\T$ for some disturbance signal $w(\cdot)$ with parameters $\btheta=(1,-1)\in\R^2$. To estimate these parameters online, we build a history stack $\mathcal{H}(t)$ by periodically sampling the signals $y(t)$ and $\bphi(t)$  and storing\footnote{Here, the size of $\mathcal{H}$ is not fixed and grows as time progresses.} the values in $\mathcal{H}(t)$. The data in $\mathcal{H}(t)$ is then used to define the vector $\bY(t)$ and matrix $\bPhi(t)$ as in \eqref{eq:Y-Phi}, which are used to estimate the parameters using the update laws in \eqref{eq:theta-hat-dot} and \eqref{eq:Gamma-dot}. The evolution of the resulting parameter estimation errors for different values of $\beta$ and disturbance signals $w(t)$ are illustrated by the solid lines in Fig. \ref{fig:simple-example}a. The value of $\gamma$ satisfying the bound in \eqref{eq:FE} for each $t\geq0$ is displayed in Fig. \ref{fig:simple-example}b, indicating that the history stack satisfies the FE condition. As guaranteed by Lemma \ref{lemma:theta-tilde} and Proposition \ref{proposition:theta-tilde}, the parameter estimation error exponentially converges to zero in the absence of disturbances (blue) and converges to a neighborhood of zero in the presence of disturbances (red and purple). Note that these signals also satisfy the PE condition, and the evolution under the traditional RLS algorithm (i.e., without the history stack \cite[Ch. 4]{IoannouSun}) is illustrated by the dashed lines in Fig. \ref{fig:simple-example}a. Although the estimation error converges to zero without the history stack, the convergence is slower, the residual bounds are larger, and such bounds are more sensitive to $\beta$.
    \end{example}

    \section{Recursive Set-Membership Identification}\label{sec:UQ}
    \begin{figure*}
        \centering
        \vspace{0.2cm}
        \includegraphics{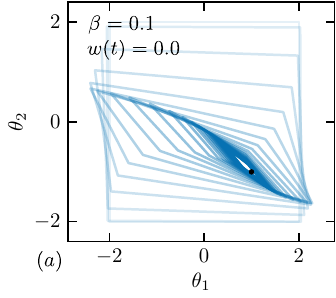}
        \hfill
        \includegraphics{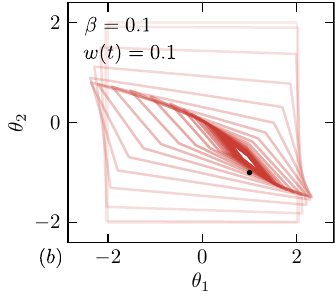}
        \hfill
        \includegraphics{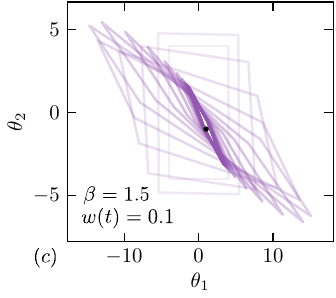}
        \vspace{-0.4cm}
        \caption{Zonotopes generated by the RLS algorithm from Example \ref{ex:zono}. In each plot, the black dot denotes $\btheta$ and the change in opacity of the zonotopes represents the evolution of time, where more transparent shapes denote zonotopes at earlier times and less transparent shapes denote zonotopes at later times. In each simulation, the update laws are defined by a history stack with $M=20$ entries, which is filled by periodically storing values of $y(t)$ and $\bphi(t)$ until the stack is full.}
        \label{fig:zono-example}
        \vspace{-0.6cm}
    \end{figure*}
\begin{figure}
        \centering
        \vspace{0.2cm}
        \includegraphics{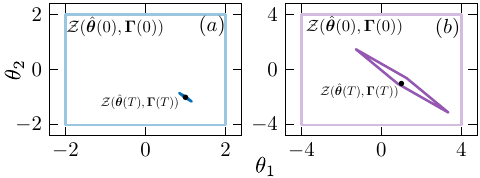}
        \vspace{-0.4cm}
        \caption{Initial and final zonotopes ($T=60$) corresponding to the different cases from Example \ref{ex:zono}. Here, Fig. \ref{fig:zono-final}(a) corresponds to the zonotopes from Fig. \ref{fig:zono-example}(a) and Fig. \ref{fig:zono-final}(b) corresponds to the zonotopes from Fig. \ref{fig:zono-example}(c).}
        \vspace{-0.6cm}
        \label{fig:zono-final}
    \end{figure}

    In this section we demonstrate how the estimation scheme introduced in the previous section naturally extends to quantifying the uncertainty in the parameter estimate $\bhat{\theta}$ through the matrix $\bGamma$. This is accomplished by blending classical parameter estimation techniques from adaptive control with set membership identification (SMID) techniques. Ultimately, this unification leads to the ability to efficiently propagate a set of all possible parameter values using the update laws proposed in the previous section. To facilitate this approach, we assume that the set of all possible parameters can be represented as a \emph{zonotope}, which are widely used for reachability analysis \cite{GirardHSCC05,GirardAR21}.
    \begin{definition}
        A set $\mathcal{Z}\subset\R^p$ is said to be a \emph{zonotope} if:
        \begin{equation}
            \mathcal{Z} = \Big\{\btheta\in\R^p\,:\,\btheta=\bc + \sum_{i=1}^q\lambda_i\bG_i,\,-1\leq\lambda_i\leq1 \Big\},
        \end{equation}
        where $\bc\in\R^p$ is the zonotope's \emph{center} and $\bG_i\in\R^p$, $i\in\{1,\dots,q\}$, are its \emph{generators}, which form a \emph{generator matrix} $\bG=\hcat(\bG_1,\dots,\bG_q)$.
    \end{definition}
    To emphasize that a zonotope depends on $\bc$ and $\bG$, we write $\mathcal{Z}(\bc,\bG)\subset\R^p$ to denote a zonotope $\mathcal{Z}$ with center $\bc$ and generator matrix $\bG$.
    The main property of zonotopes we exploit is that they are closed under affine transformation:
    \begin{equation}
        \bA \mathcal{Z}(\bc,\bG) + b = \mathcal{Z}(\bA\bc + \bb, \bA\bG),
    \end{equation}
    for a matrix $\bA$ and vector $\bb$ of appropriate dimensions. The primary observation that facilitates our extension of the previous estimation scheme to set-based estimates is that the point-wise estimate $\bhat{\theta}(t)$ from \eqref{eq:theta-hat} at any given point in time is simply an affine transformation of the initial condition:
    \begin{equation*}
        \begin{aligned}
            \bhat{\theta}(t) = & \underbrace{e^{-\beta t}\bGamma(t)\bP}_{\bA(t)}\bhat{\theta}_0 + \underbrace{\bGamma(t)\int_{0}^{t}e^{-\beta(t-s)}\bPhi(s)\T \bY(s)\mathrm{d}s}_{\bb(t)}.
        \end{aligned}
    \end{equation*}
    Replacing $\bhat{\theta}_0$ with $\mathcal{Z}(\bhat{\theta}_0,\bP^{-1})$ then yields:
    \begin{equation}\label{eq:zonotope-affine-transformation}
        \begin{aligned}
            \bA(t)\mathcal{Z}(\bhat{\theta}_0,\bP^{-1}) + \bb(t) = & \mathcal{Z}(\bA(t)\bhat{\theta}_0 + \bb(t), \bA(t)\bP^{-1}) \\ 
            = & \mathcal{Z}(\bhat{\theta}(t), \bGamma(t)).
        \end{aligned}
    \end{equation}
    Ultimately, we would like to show that $\btheta\in\mathcal{Z}(\bhat{\theta}(t), \bGamma(t))$ for each $t$, which allows for quantifying the uncertainty in a given parameter estimate $\bhat{\theta}(t)$ via $\mathcal{Z}(\bhat{\theta}(t), \bGamma(t))$.
    Further results are facilitated by the following lemma.
    \begin{lemma}[\cite{SadraCDC19}]\label{lemma:zonotope-containment}
        Consider zonotopes $\mathcal{Z}(\bc_1,\bG_1)\subset\R^p$ and $\mathcal{Z}(\bc_2,\bG_2)\subset\R^p$ with centers $\bc_1,\bc_2\in\R^p$ and generators $\bG_1\in\R^{p\times q}$, $\bG_2\in\R^{p\times r}$. Then, $\mathcal{Z}(\bc_1,\bG_1)\subseteq\mathcal{Z}(\bc_2,\bG_2)$ if there exists a pair $(\bz,\bQ)\in\R^p\times \R^{r\times q}$ such that 
        \begin{subequations}\label{eq:zono-contain}
            \begin{equation}\label{eq:zono-contain-1}
                \bG_1 = \bG_2\bQ
            \end{equation}
            \begin{equation}\label{eq:zono-contain-2}
                \bc_2 - \bc_1 = \bG_2\bz
            \end{equation}
            \begin{equation}\label{eq:zono-contain-3}
                \|\hcat(\bQ,\bz)\|_{\infty}\leq 1.
            \end{equation}
        \end{subequations}
    \end{lemma}
    The following theorem constitutes the first main result of this paper, providing conditions under which $\btheta$ is guaranteed to remain in the zonotopes generated by \eqref{eq:zonotope-affine-transformation}.
    \begin{theorem}\label{theorem:zonotope}
        Let the conditions of Lemma \ref{lemma:theta-tilde} hold and define
        \begin{equation}
            \bar{w}_{\infty}\coloneqq\sup_{t\geq0}\|\bPhi(t)\T\bW(t)\|_{\infty}.
        \end{equation}
        Furthermore, suppose that:
        \begin{equation}\label{eq:zono-contain-cond}
            \|\bP(\bhat{\theta}_0 - \btheta)\|_{\infty} + \frac{\bar{w}_{\infty}}{\beta} \leq 1.
        \end{equation}
        Then:
        \begin{equation}\label{eq:theta-zono-contain}
            \btheta\in \mathcal{Z}(\bhat{\theta}(t), \bGamma(t)),\quad \forall t\geq0.
        \end{equation}
    \end{theorem}
    \begin{proof}
        We check that Lemma \ref{lemma:zonotope-containment} holds for:
        \begin{equation*}
            \{\btheta\}=\mathcal{Z}(\btheta,\mathbf{0}) \subseteq \mathcal{Z}(\bhat{\theta}(t),\bGamma(t)),
        \end{equation*}
        for each $t\geq0$. Noting that $\bG_1=\mathbf{0}$ and $\bG_2=\bGamma$, we find that \eqref{eq:zono-contain-1} is satisfied by $\bQ=\bzero$ since $\bGamma(t)$ is positive definite for each $t\geq0$. Noting that $\bc_1=\btheta$ and $\bc_2=\bhat{\theta}$, we find that \eqref{eq:zono-contain-2} is satisfied by:
        \begin{equation*}
            \bz = \bGamma(t)^{-1}(\bhat{\theta}(t) - \btheta) = -\bGamma(t)^{-1}\btilde{\theta}(t).
        \end{equation*}
        To complete the proof, we must show that $\bQ$ and $\bz$ satisfy \eqref{eq:zono-contain-3}. We first note that since $\bQ=\bzero$ and the induced $\infty$-norm for matrices is equal to the absolute row sum, we have:
        \begin{equation*}
            \|\hcat(\bQ,\bz)\|_{\infty} = \|\bz\|_{\infty} = \|-\bGamma^{-1}(t) \btilde{\theta}(t)\|.
        \end{equation*}
        It then follows from the above and Lemma \ref{lemma:theta-tilde} that:
        \begin{equation*}
            \begin{aligned}
                \|\bz\|_{\infty} = \left\Vert e^{-\beta t}\bP(\bhat{\theta}_0 - \btheta) +\int_{0}^{t}e^{-\beta(t-s)}\bPhi(s)\T\bW(s)\mathrm{d}s \right\Vert_{\infty}
            \end{aligned}
        \end{equation*}
        which, after using the triangle inequality and the homogeneity of norms, implies that:
        \begin{equation*}
            \begin{aligned}
                \|\bz\|_{\infty} \leq & e^{-\beta t} \|\bP(\bhat{\theta}_0 - \btheta)\|_{\infty} \\
               & + \left\Vert \int_{0}^{t}e^{-\beta(t-s)}\bPhi(s)\T\bW(s)\mathrm{d}s \right\Vert_{\infty} \\ 
               \leq & \|\bP(\bhat{\theta}_0 - \btheta)\|_{\infty} + \left\Vert \int_{0}^{t}e^{-\beta(t-s)}\bPhi(s)\T\bW(s)\mathrm{d}s \right\Vert_{\infty},
            \end{aligned}
        \end{equation*}
        where the second inequality follows from $e^{-\beta t}\leq 1$. Using a similar argument as in the proof of Proposition \ref{proposition:theta-tilde}, we may bound the second term in the above inequality as:
        \begin{equation*}
            \left\Vert \int_{0}^{t}e^{-\beta(t-s)}\bPhi(s)\T\bW(s)\mathrm{d}s \right\Vert_{\infty} \leq \frac{\bar{w}_{\infty}}{\beta},
        \end{equation*}
        which, after using \eqref{eq:zono-contain-cond}, implies that:
        \begin{equation*}
            \|\bz\|_{\infty} \leq \|\bP(\bhat{\theta}_0 - \btheta)\|_{\infty} + \frac{\bar{w}_\infty}{\beta} \leq 1.
        \end{equation*}
        It then follows that the conditions of Lemma \ref{lemma:zonotope-containment} are satisfied, which implies that \eqref{eq:theta-zono-contain} holds, as desired.
    \end{proof}
    The preceding theorem provides the machinery to construct a zonotope at each instant in time that is guaranteed to contain the true parameters. The pair $(\bhat{\theta}_0,\bP)$ from \eqref{eq:LS-cost} captures prior knowledge of the parameters in that $\bhat{\theta}_0$ corresponds to the best initial guess of the parameters and $\bP$ represents the confidence of such a guess. The condition in \eqref{eq:zono-contain-cond} indicates that $\bP$ must be chosen small enough for a given level of initial parameter estimation error $\|\btheta - \bhat{\theta}_0\|_{\infty}$ and the discount factor $\beta$ must be chosen large enough for a given level of disturbance affecting \eqref{eq:LRE} so that $\mathcal{Z}(\bhat{\btheta}(t),\bGamma(t))$ contains $\btheta$ for all $t\geq0$. 
    \begin{example}\label{ex:zono}
        We continue Example \ref{example:simple-example} to demonstrate how the approach in this section is used to construct a set-based estimate of $\btheta$. We begin with the disturbance-free case, which, by Theorem \ref{theorem:zonotope}, requires selecting $\bP$ such that $\|\bP(\bhat{\theta}_0 - \btheta)\|_{\infty}\leq1$. We take $\bhat{\theta}_0=\bzero$ and assume prior bounds on $\btheta$ in the sense that $\|\bhat{\theta}_0 - \btheta\|\leq 2$, allowing to take $\bP=\tfrac{1}{2}\bI$ to satisfy \eqref{eq:zono-contain-cond}. The initial estimates are then used to define the initial conditions of the ODEs from \eqref{eq:theta-hat-dot} and \eqref{eq:Gamma-dot} as $\bhat{\theta}(0)=\bhat{\theta}_0$ and $\bGamma(0)=\bP^{-1}$, respectively, the solutions of which $t\mapsto\bhat{\theta}(t),\bGamma(t)$ are used to construct a zonotope $\mathcal{Z}(\bhat{\theta}(t),\bGamma(t))$ whose evolution is illustrated in Fig. \ref{fig:zono-example}a. As guaranteed by Theorem \ref{theorem:zonotope}, these zonotopes contain $\btheta$ for all time. We now introduce a disturbance $w(t)=0.1$ and propagate the ODEs in \eqref{eq:theta-hat-dot} and \eqref{eq:Gamma-dot} with the same initial conditions and discount factor $\beta$ as in the disturbance-free case. The resulting zonotopes are illustrated Fig. \ref{fig:zono-example}b, where, after a certain period of time, the zonotopes no longer contain $\btheta$ as the condition in \eqref{eq:zono-contain-cond} is not satisfied. To remedy this one must decrease $\bP$ and increase $\beta$ commensurate with the value of $\bar{w}_{\infty}$. To satisfy \eqref{eq:zono-contain-cond} we take $\bP = \tfrac{1}{4}\bI$ and $\beta=1.5$ and propagate the ODEs in \eqref{eq:theta-hat-dot} and \eqref{eq:Gamma-dot}, the results of which are shown in Fig. \ref{fig:zono-example}(c). With these modified values of $\bP$ and $\beta$, the zonotopes in Fig. \ref{fig:zono-example}c contain the true parameters at all times, but are more conservative. Although the intermediate zonotopes are more conservative than in the disturbance-free case, the zonotopes generated by such an approach for longer time horizons significantly reduce the conservatism of the initial zonotope (cf. Fig. \ref{fig:zono-final}).
    \end{example}

    \begin{remark}\label{remark:polytope}
        Although Theorem \ref{theorem:zonotope} guarantees that $\btheta\in\mathcal{Z}(\bhat{\theta}(t),\bGamma(t))$ for all $t\in\R_{\geq0}$, this does not necessarily imply that $\mathcal{Z}(\bhat{\theta}(t),\bGamma(t))\subseteq \mathcal{Z}(\bhat{\theta}(0),\bGamma(0))$ for all $t\in\R_{\geq0}$, as illustrated in Fig. \ref{fig:zono-example}. To remedy this, one may define:
        \begin{equation*}
            \mathcal{P}(t) = \mathcal{Z}(\bhat{\theta}(0),\bGamma(0)) \cap \mathcal{Z}(\bhat{\theta}(t),\bGamma(t)),
        \end{equation*}
        which produces a polytope (not necessarily a zonotope) that satisfies $\btheta\in\mathcal{P}(t)$ for all $t\in\R_{\geq0}$.
    \end{remark}

    \section{Robust Adaptive Safety-Critical Control}\label{sec:cbf}
    We now illustrate how the techniques from the previous section can be used to develop safety-critical controllers for systems with parametric uncertainty. We consider nonlinear control systems with parametric uncertainty of the form:
    \begin{equation}\label{eq:dyn}
        \dot{\bx} = \bf(\bx) + \bF(\bx)\btheta + \bg(\bx)\bu + \bd,
    \end{equation}
    where $\bx\in\R^n$ is the state, $\bu\in\R^m$ is the control, $\bd\in\R^n$ is a disturbance, and $\bf\,:\,\R^n\rightarrow\R^n$, $\bg\,:\,\R^n\rightarrow\R^{n\times m}$, and $\bF\,:\,\R^n\rightarrow\R^{n\times p}$ are locally Lipschitz continuous. Hence, given a locally Lipschitz feedback controller $\bu=\bk(\bx,\bhat{\theta},\bGamma)$, a piecewise continuous disturbance signal $t\mapsto\bd(t)$, and an initial condition $(\bx_0,\bhat{\theta}_0,\bGamma_0)\in\R^n\times\R^p\times\R^{p\times p}$, there exist trajectories $t\mapsto\bx(t),\bhat{\theta}(t),\bGamma(t)$ of the closed-loop system:
    \begin{equation}\label{eq:closed-loop-dyn}
    \begin{aligned}
        \dot{\bx} = & \bf(\bx) + \bF(\bx)\btheta + \bg(\bx)\bk(\bx,\bhat{\theta},\bGamma) + \bd(t), \\ 
        \bdothat{\theta} = & \bGamma\bPhi(t)\T\big[\bY(t) - \bPhi(t)\bhat{\theta}\big], \\
        \dot{\bGamma} = & \beta\bGamma -\bGamma \bPhi(t)\T\bPhi(t)\bGamma,
    \end{aligned}
    \end{equation}
    defined on some maximal interval of existence $I\subseteq\R_{\geq0}$, which we take to be the nonnegative real line $I=\R_{\geq0}$ for ease of exposition. Our main objective in this section is to design feedback controllers for \eqref{eq:dyn} such that the closed-loop system \eqref{eq:closed-loop-dyn} is \emph{safe}, a concept often formalized using the framework of set invariance. Formally, we say that \eqref{eq:closed-loop-dyn} is \emph{safe} on a set $\mathcal{C}\subset\R^n$ from an initial condition $(\bx_0,\bhat{\theta}_0,\bGamma_0)\in\C\times\R^p\times\R^{p\times p}$ if the resulting trajectory satisfies $\bx(t)\in\C$ for all $t\in \R_{\geq0}$. By considering sets of the form:
    \begin{equation}\label{eq:C}
        \C = \{\bx\in\R^n\,:\,h(\bx)\geq0\},
    \end{equation}
    for a continuously differentiable function $h\,:\,\R^n\rightarrow\R$, one can certify the safety of \eqref{eq:closed-loop-dyn} using barrier functions.
    \begin{definition}
        A continuously differentiable function $h\,:\,\R^n\rightarrow\R$ defining a set $\C\subset\R^n$ as in \eqref{eq:C} is said to be a \emph{robust adaptive barrier function} (RaBF) for \eqref{eq:closed-loop-dyn} on $\C$ if zero is a regular value of $h$ and there exists an extended class $\mathcal{K}_{\infty}$ function $\alpha$ such that for all $(\bx,\bhat{\theta},\bGamma)\in\R^n\times\R^p\times\R^{p\times p}$:
        \begin{equation}\label{eq:adaptive-barrier}
        \begin{aligned}
            L_{\bf}h(\bx) + L_{\bF}h(\bx)\btheta + L_{\bg}h(\bx)\bk(\bx,\bhat{\theta},\bGamma) \\ \geq - \alpha(h(\bx)) + \bar{d}\|\nabla h(\bx)\|_2,
        \end{aligned}
        \end{equation}
        where $\bar{d}\coloneqq \sup_{t\geq0}\|\bd(t)\|_2$.
    \end{definition}
    The following lemma shows that the existence of an RaBF is sufficient to conclude the safety of \eqref{eq:closed-loop-dyn}.
    \begin{lemma}\label{lemma:aBF}
        Let $h\,:\,\R^n\rightarrow\R$ be an RaBF for \eqref{eq:closed-loop-dyn} on a set $\C\subset\R^n$ as in \eqref{eq:C}. Then, \eqref{eq:closed-loop-dyn} is safe on $\mathcal{C}$ for each $(\bx_0,\bhat{\theta}_0,\bGamma_0)\in\C\times\R^p\times\R^{p\times p}$.
    \end{lemma}
    \begin{proof}
        The time-derivative of $h$ is bounded as:
        \begin{equation*}
            \begin{aligned}
                \dot{h} = & L_{\bf}h(\bx) + L_{\bF}h(\bx)\btheta + L_{\bg}h(\bx)\bk(\bx,\bhat{\theta},\bGamma) + \nabla h(\bx)\cdot \bd\\ 
                \geq & L_{\bf}h(\bx) + L_{\bF}h(\bx)\btheta + L_{\bg}h(\bx)\bk(\bx,\bhat{\theta},\bGamma) -\bar{d}\|\nabla h(\bx)\|_2\\ 
                \geq & -\alpha(h(\bx)),
            \end{aligned}
        \end{equation*}
        where the final inequality follows from \eqref{eq:adaptive-barrier}. Thus, since zero is a regular value of $h$ and for all $(\bx,\bhat{\theta},\bGamma,t)$, $h(\bx)=0\implies \dot{h}(\bx,\bhat{\theta},\bGamma,t)\geq0$,
        it follows from Nagumo's Theorem \cite[Ch. 4.2]{AbrahamMarsdenRatiu} that $h(\bx(t))\geq0$ for all $t\in\R_{\geq0}$, which implies $\bx(t)\in\mathcal{C}$ for all $t\in \R_{\geq0}$, as desired.
    \end{proof}
    The previous lemma demonstrates that the satisfaction of inequality \eqref{eq:adaptive-barrier} is sufficient to conclude the safety of the closed-loop system \eqref{eq:closed-loop-dyn}; however, directly verifying such an inequality presents challenges since it depends on $\btheta$, which is unknown. In what follows, we illustrate how the results on zonotopes from the previous section can be used to construct verifiable conditions for safety. To this end, note that if $\mathcal{Z}(\bc,\bG)$ is a zonotope, then each $\bz\in\mathcal{Z}(\bc,\bG)$ satisfies the element-wise inequality \cite{KasraCDC22}:
    \begin{equation}\label{eq:zonotope-bounds}
        \bc - \sum_{i=1}^q|\bG_i| \leq \bz \leq \bc + \sum_{i=1}^q|\bG_i|,
    \end{equation}
    where $|\bG_i|$ denotes element-wise absolute value. The above property is now used to establish bounds on $L_{\bF}h(\bx)\btheta$.
    \begin{lemma}\label{lemma:zono-bounds}
        Suppose that $\btheta\in\mathcal{Z}(\bhat{\theta},\bGamma)\subset\R^p$. Then:
        \begin{equation}\label{eq:Lie-derivative-bound}
            L_{\bF}h(\bx)\btheta \geq L_{\bF}h(\bx)\bhat{\theta} - \|L_{\bF}h(\bx)\bGamma\|_1,\quad \forall \bx\in\R^n.
        \end{equation}  
    \end{lemma}
    \begin{proof}
        Note than since $\{\btheta\}=\mathcal{Z}(\btheta,0)\subseteq \mathcal{Z}(\bhat{\theta},\bGamma)$, we have:
        \begin{equation*}
        \begin{aligned}
            \mathcal{Z}(L_{\bF}h(\bx)\btheta,0) \subseteq \mathcal{Z}(L_{\bF}h(\bx)\bhat{\theta}, L_{\bF}h(\bx)\bGamma)\subseteq\R.
        \end{aligned}
        \end{equation*}
        It then follows from \eqref{eq:zonotope-bounds} that: 
        \begin{equation}\label{eq:zono-bound-Lie-derivative}
        \begin{aligned}
            z \geq & L_{\bF}h(\bx)\bhat{\theta} - \| L_{\bF}h(\bx)\bGamma\|_1,
        \end{aligned}
        \end{equation}
        for each $z\in \mathcal{Z}(L_{\bF}h(\bx)\bhat{\theta}, L_{\bF}h(\bx)\bGamma)$.
        Since: $$L_{\bF}h(\bx)\btheta\in \mathcal{Z}(L_{\bF}h(\bx)\bhat{\theta}, L_{\bF}h(\bx)\bGamma),$$ \eqref{eq:zono-bound-Lie-derivative} implies \eqref{eq:Lie-derivative-bound}, as desired.
    \end{proof}

    We now have all the tools in place to introduce a controlled version of a RaBF, which facilitates the synthesis of controllers enforcing safety of the closed-loop system \eqref{eq:closed-loop-dyn}.

    \begin{definition}
        A continuously differentiable function $h\,:\,\R^n\rightarrow\R$ defining a set $\C\subset\R$ as in \eqref{eq:C} is said to be a \emph{robust adaptive control barrier function} (RaCBF) for \eqref{eq:dyn} on $\C$ if for all $(\bx,\bhat{\theta},\bGamma)\in\R^n\times\R^p\times\R^{p\times p}$:
        \begin{equation}\label{eq:RaCBF}
            \begin{aligned}
                \sup_{\bu\in\R^m} & \Big\{L_{\bf}h(\bx) + L_{\bF}h(\bx)\bhat{\theta} + L_{\bg}h(\bx)\bu \Big\} \\ & \geq - \alpha(h(\bx)) + \|L_{\bF}h(\bx)\bGamma \|_{1} + \bar{d}\|\nabla h(\bx)\|_2.
            \end{aligned}
        \end{equation}
    \end{definition}
    A RaCBF allows one to construct the set-valued map:
    \begin{equation}\label{eq:K-cbf}
        \begin{aligned}
            \bK(\bx,\bhat{\theta},\bGamma) \coloneqq  \Big\{\bu\in\R^m\,:\, L_{\bf}h(\bx) + L_{\bF}h(\bx)\bhat{\theta} + L_{\bg}h(\bx)\bu \\ 
             \geq - \alpha(h(\bx)) + \|L_{\bF}h(\bx)\bGamma \|_{1} + \bar{d}\|\nabla h(\bx)\|_2\Big\},
        \end{aligned}
    \end{equation}
    assigning to each triple $(\bx,\bhat{\theta},\bGamma)$ the set of all control values satisfying the condition in \eqref{eq:RaCBF}. The following theorem shows that when $\btheta\in\mathcal{Z}(\bhat{\theta}(t),\bGamma(t))$ for all $t\in\R_{\geq0}$, any locally Lipschitz controller drawn from \eqref{eq:K-cbf} renders the closed-loop system safe.
    \begin{theorem}
        Let $h\,:\,\R^n\rightarrow\R$ be a RaCBF for \eqref{eq:dyn} on a set $\mathcal{C}\subset\R^n$ as in \eqref{eq:C}. Let $\bk\,:\,\R^n\times\R^p\times\R^{p\times p}\rightarrow\R^m$ be a locally Lipschitz controller satisfying $\bk(\bx,\bhat{\theta},\bGamma)\in \bK(\bx,\bhat{\theta},\bGamma)$ for all $(\bx,\bhat{\theta},\bGamma)\in\R^n\times\R^p\times\R^{p\times p}$, and let $t\mapsto \bx(t),\bhat{\theta}(t),\bGamma(t)$ be the trajectories of the resulting closed-loop system \eqref{eq:closed-loop-dyn} from a given initial condition $(\bx_0,\bhat{\theta}_0,\bGamma_0)\in\C\times\R^p\times\R^{p\times p}$. Then, if $\btheta\in\mathcal{Z}(\bhat{\theta}(t),\bGamma(t))$ for all $t\in\R_{\geq0}$, the closed-loop system is safe on $\mathcal{C}$.
    \end{theorem}
    \begin{proof}
        Taking the derivative of $h$ along the closed-loop system trajectory and lower-bounding (omitting time-dependency for ease of presentation) yields:
        \begin{equation*}
            \begin{aligned}
                \dot{h} = & L_{\bf}h(\bx) + L_{\bF}h(\bx)\btheta + L_{\bg}h(\bx)\bk(\bx,\bhat{\theta},\bGamma) + \nabla h(\bx)\cdot \bd \\
                \geq & L_{\bf}h(\bx) + L_{\bF}h(\bx)\btheta + L_{\bg}h(\bx)\bk(\bx,\bhat{\theta},\bGamma) -\bar{d} \|\nabla h(\bx)\|_2 \\
                \geq & L_{\bf}h(\bx) + L_{\bF}h(\bx)\bhat{\theta} + L_{\bg}h(\bx)\bk(\bx,\bhat{\theta},\bGamma) \\
                & -\|L_{\bF}h(\bx)\bGamma\|_1 -\bar{d} \|\nabla h(\bx)\|_2 \\
                \geq & - \alpha(h(\bx)),
            \end{aligned}
        \end{equation*}
        where the second inequality follows from Lemma \ref{lemma:zono-bounds} since $\btheta\in\mathcal{Z}(\bhat{\theta}(t),\bGamma(t))$ for all $t\geq0$, and the third from \eqref{eq:K-cbf}. The remainder of the proof follows that of Lemma \ref{lemma:aBF}.
    \end{proof}

\begin{figure}
        \centering
        \vspace{0.2cm}
        \includegraphics{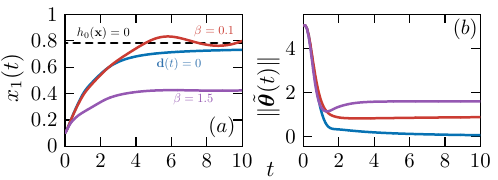}
        \vspace{-0.4cm}
        \caption{Trace of the pendulum's position (a) and resulting parameter estimation error (b) for the simulations from Example \ref{ex:cbf}.}
        \vspace{-0.6cm}
        \label{fig:cbf-ex}
    \end{figure}

    Given a RaCBF, controllers enforcing safety of the closed-loop system can be synthesized via the quadratic program:
    \begin{equation}\label{eq:safety-filter}
        \begin{aligned}
            \min_{\bu\in\R^m} & \quad \tfrac{1}{2}\|\bu - \bk_{\mathrm{d}}(\bx,\bhat{\theta},\bGamma))\|_2^2 \\
            \mathrm{s.t.} & \quad L_{\bf}h(\bx) + L_{\bF}h(\bx)\bhat{\theta} + L_{\bg}h(\bx)\bu \\ & \geq -\alpha(h(\bx)) +\|L_{\bF}h(\bx)\bGamma\|_1 +\bar{d} \|\nabla h(\bx)\|_2,
        \end{aligned}
    \end{equation}
    where $\bk_{\mathrm{d}}\,:\,\R^n\times\R^p\times\R^{p\times p}\rightarrow\R^m$ is a nominal feedback controller. Note that safety of the closed-loop system under such a controller is contingent on the assumption that $\btheta\in\mathcal{Z}(\bhat{\theta}(t),\bGamma(t))$ for all $t\in\R_{\geq0}$, which holds if the update laws in \eqref{eq:theta-hat-dot} and \eqref{eq:Gamma-dot} satisfy the conditions of Theorem \ref{theorem:zonotope}.
    \begin{example}\label{ex:cbf}
        Consider an inverted pendulum with dynamics:
        \begin{equation*}
            \underbrace{
            \begin{bmatrix}
                \dot{x}_1 \\ \dot{x}_2
            \end{bmatrix}}_{\dot{\bx}}
            =
            \underbrace{
            \begin{bmatrix}
                x_2 \\ 0
            \end{bmatrix}
            }_{\bf(\bx)}
            +
            \underbrace{
            \begin{bmatrix}
                0 & 0 \\ \sin(x_1) & -x_2
            \end{bmatrix}}_{\bF(\bx)}
            \underbrace{
            \begin{bmatrix}
                \theta_1 \\ \theta_2
            \end{bmatrix}}_{\btheta}
            +
            \underbrace{
            \begin{bmatrix}
                0 \\ 1
            \end{bmatrix}}_{\bg(\bx)}
            \bu + \bd(t)
        \end{equation*}
        with state $\bx=(x_1,x_2)$ representing angular position and velocity, input $\bu\in\R$ representing torque applied to the base of the pendulum, and $\btheta=(\theta_1,\theta_2)\in\R^2$ representing uncertain parameters associated with gravitational effects and damping effects, respectively. The objective is to design a controller such that the orientation of pendulum $x_1$ satisfies $x_1\in[-\tfrac{\pi}{4},\tfrac{\pi}{4}]$, leading to the safety constraint $h_0(\bx) = (\tfrac{\pi}{4})^2 - x_1^2\geq0,$ which is used to construct a CBF using barrier backstepping \cite{AndrewCDC22}. As the dynamics are subject to uncertain parameters and additive disturbances, we leverage the RaCBF approach developed in this section to design such a controller. To estimate the parameters online we leverage the approach from Sec. \ref{sec:RLS} using a history stack with $M=20$ entries, where\footnote{Here, we simply take $y=\dot{\bx} - \bf(\bx) - \bg(\bx)\bu$ and $\bphi=\bF(\bx)$, implying $w=\bd$, where $\bd$ may also capture noise present in measurements/estimates of $\dot{\bx}$.} \eqref{eq:LRE} is sampled every 0.1 seconds until the stack is full. We take our initial parameter estimate as $\bhat{\theta}_0=(5,2)$, reflecting the assumption that weaker gravitational effects and more damping are present, and choose $\bP=\mathrm{diag}(0.1, 0.4)$ to reflect our confidence in such an estimate. With such initial conditions, we take $\beta=0.1$ and simulate the closed-loop adaptive system \eqref{eq:closed-loop-dyn} under the influence of the controller in \eqref{eq:safety-filter}, where $\bk_{\mathrm{d}}(\bx)=0$. The system is simulated from such initial conditions for two different scenarios: i) without disturbances $\bd(t)=0$; ii) with disturbance $\bd(t)=[0.0, -0.4\sin(t)]\T$ but where the disturbance is not accounted for in the controller (i.e, $\bar{d}=0$ in \eqref{eq:safety-filter}) or update law (i.e., via $\bP$ and $\beta$), the results of which are shown in Fig. \ref{fig:cbf-ex} by the blue and red curves, respectively. In the disturbance-free case, safety is maintained and the parameter estimation error converges to zero; when disturbances are present and not accounted for safety is violated. To enforce safety in the presence of disturbances, we rerun the system with $\beta=1.5$ and $\bar{d}=0.4$, the results of which are shown by the purple curves in Fig. \ref{fig:cbf-ex}, leading to a safe closed-loop system, although in a somewhat conservative\footnote{This conservatism can be mitigated using the polytopic approach from Remark \ref{remark:polytope} at the cost of additional computations.} fashion.
    \end{example}

    \section{Discussion and Conclusions}
    In this paper, we presented a framework for online parameter estimation and uncertainty quantification that unites tools from set membership identification and concurrent learning adaptive control. Such developments were motivated by their eventual integration with control barrier functions, which ultimately allowed for synthesizing controllers enforcing safety of nonlinear systems with parametric uncertainties and additive disturbances. We believe these results may serve as the starting point for various extensions, such as integration with model predictive control, leveraging other set representations for uncertainty quantification, and considering stochastic disturbances. 

    \bibliographystyle{ieeetr}
    \bibliography{
        biblio/adaptive,
        biblio/books,
        biblio/barrier,
        biblio/hybrid,
        biblio/mpc,
        biblio/learning,
        biblio/nonlinear}
    
\end{document}